\documentclass[twocolumn,pre,twoside,showpacs,preprintnumbers,floatfix]{revtex4}

\usepackage{amsmath}
\usepackage{amssymb}
\usepackage{graphicx}
\usepackage{dcolumn}
\usepackage{bm}

\def\m#1{\mathrm{#1}}
\def\Eq#1{(\ref{eq:#1})}

\def\epsilon{\varepsilon}
\def\theta{\vartheta}
\def\rho{\varrho}
\def\Int#1#2#3{\int\limits_{#1}^{#2}\!\mathrm{d}{#3}\;}

\def\dps{\displaystyle}
\def\sign{\mathop{\mathrm{sign}}}

\begin{document}


\title{Curvature dependence of the electrolytic liquid-liquid interfacial tension}

\author{Markus Bier}
\email{m.bier@uu.nl}

\author{Joost de Graaf}
\author{Jos Zwanikken}
\author{Ren\'e van Roij}

\affiliation
{
   Institute for Theoretical Physics, 
   Utrecht University, 
   Leuvenlaan 4, 
   3584\,CE Utrecht, 
   The Netherlands
}

\date{17 September 2008}

\begin{abstract}
   The interfacial tension of a liquid droplet surrounded by another liquid in the presence of microscopic ions
   is studied as a function of the droplet radius. 
   An analytical expression for the interfacial tension is obtained within a linear Poisson-Boltzmann theory
   and compared with numerical results from non-linear Poisson-Boltzmann theory.
   The excess liquid-liquid interfacial tension with respect to the pure, salt-free liquid-liquid interfacial 
   tension is found to decompose into a curvature-independent part due to short-ranged interfacial effects and
   a curvature-dependent electrostatic contribution.
   Several curvature-dependent regimes of different scaling of the electrostatic excess interfacial tension
   are identified.
   Symmetry relations of the interfacial tension upon swapping droplet and bulk liquid are found to hold in the 
   low-curvature limit, which, e.g., lead to a sign change of the excess Tolman length.
   For some systems a low-curvature expansion up to second order turns out to be applicable if and only if the 
   droplet size exceeds the Debye screening length in the droplet, independent of the Debye length in the bulk.
\end{abstract}


\maketitle


\section{Introduction}

Common wisdom in emulsion science tells that, in order to kinetically stabilize an emulsion of water and oil,
say, surfactants are needed in order to decrease the interfacial tension thereby decreasing the thermodynamic
force causing droplet coalescence \cite{Safran2003}.
This picture has been upset by Leunissen et al.\ who showed experimentally that in certain additive-free
water-oil mixtures micron-sized water droplets in oil may be stabilized electrostatically by absorbing ions 
present in the system \cite{Leunissen2007a,Leunissen2007b}. 
Several aspects of these experiments such as the proposed charging of the water droplets due to an unequal 
partitioning \cite{Zwanikken2007,Bier2008,Zwanikken2008} and the formation of a colloidal crystal of water droplets 
\cite{deGraaf2008} can be understood theoretically within a simple Poisson-Boltzmann model.
However, the rather unimodal size distribution of the water droplets in the above-mentioned experiments has not
been explained so far.
A similar observation has been made by Sacanna et al.\ who found experimental indications of the existence of 
thermodynamically favored droplet radii in certain emulsions stabilized by nano-sized colloids \cite{Sacanna2007}.
A thermodynamically favored droplet radius requires a radius dependent water-oil interfacial tension because 
otherwise the global minimum of the free energy would be attained for one single macroscopic drop.
One is thereby led to the problem of analyzing the liquid-liquid interfacial tension as a function of the droplet
radius.

The study of the curvature dependence of liquid-vapor surface tensions has been pioneered
by Gibbs \cite{Gibbs1961}, Tolman \cite{Tolman1949}, and Kirkwood and Buff \cite{Kirkwood1949}.
Tolman introduced a low-curvature expansion of the form 
$\gamma(a)/\gamma(\infty)\simeq 1/(1+2\delta/a)\simeq 1-2\delta/a$ where $a$ denotes the radius of curvature,
$\gamma(a)$ is the surface tension of the curved surface, and $\gamma(\infty)$ is its planar value.
The parameter $\delta$, which has the dimension of length, is called the \emph{Tolman length} and it can be 
identified with the spatial distance between the Gibbs dividing surface and the surface of tension.
In the last decades the concept of a curvature dependent liquid-vapor surface tension has been taken up within
various studies on critical phenomena \cite{Fisher1984}, interface elasticity \cite{Blokhuis1992}, and nucleation
\cite{Talanquer1995,Barrett1999}.

However, whereas in all these investigations the droplet and the surrounding bulk were composed of the same
substance, albeit in different phases, here a mixture of two different liquids and ions is studied. 
Moreover, only the excess interfacial tension due to the electrolyte is of interest here while the two liquids
forming droplet and bulk merely act as external fields onto the ions.

The present investigation is carried out within the spherical version of the model studied in Ref.~\cite{Bier2008} 
(Sec.~\ref{sec:model}).
As in Ref.~\cite{Bier2008} linearization of the Poisson-Boltzmann equation offers the possibility of closed 
analytical expressions for the interfacial tension (Sec.~\ref{sec:linear}).
In Sec.~\ref{sec:discuss} the approximative analytical expressions for the interfacial tension will be shown 
to at least qualitatively, in many realistic cases even quantitatively, agree with the numerical results 
obtained within the full, non-linear theory.
The main conclusion will be that it is precisely the electrostatic contribution to the interfacial tension that 
brings about a curvature dependence, which, however, is usually insignificant to serve as an explanation for 
unimodal radius distributions in the emulsions by Leunissen et al.\ mentioned above (Sec.~\ref{sec:conclusions}).
On the other hand, the curvature-dependent electrostatic contribution to the interfacial tension can be 
expected to increase considerably in magnitude if highly charged colloids instead of monovalent ions are present.
Under these conditions, however, the approximations made in the present work are not a priori justified, and it
is left for future studies to investigate the influence of valency on the qualitative picture to be drawn
here, which corresponds to the low-valency limit. 


\section{\label{sec:model}Model and Formalism}

In the following dimensionless quantities are expessed in units of the thermal energy $k_\m{B}T$, the elementary 
charge $e$, and the vacuum Bjerrum length $\dps\ell=\frac{e^2}{4\pi\epsilon_\m{vac}k_\m{B}T}$ with the 
permeability of the vacuum $\epsilon_\m{vac}$.
Dimensionful quantities are denoted by the same symbol as the corresponding dimensionless quantities.

Consider a liquid spherical droplet of radius $a$ and relative dielectric constant $\epsilon_\m{d}$ surrounded 
by bulk liquid of relative dielectric constant $\epsilon_\m{b}$. 
Due to the spherical symmetry of the setting the only relevant positional variable is the distance 
$r\in[0,\infty)$ from the droplet center.
Monovalent cations ($+$ ions) and anions ($-$ ions) are distributed in both liquids.
The difference in solvation free energy of a $\pm$ ion in the droplet with respect to the
bulk liquid is denoted by $f_\pm$, which, within the Born approximation \cite{Born1920}, can be estimated by 
$\dps f_\pm=\frac{1}{2a_\pm}\Big(\frac{1}{\epsilon_\m{d}}-\frac{1}{\epsilon_\m{b}}\Big)$ with the ion radius 
$a_\pm$.
As in Ref.~\cite{Bier2008} all interfacial effects due to, e.g., smooth interfaces, finite ion size, van der
Waals forces, and image charges, which are short ranged as compared to the electrostatic potential, are
accounted for by introducing solvent-induced ion potentials $V_\pm(r)=f_\pm\Theta(a+s-r)$ with $\Theta$ the
Heaviside function.
Note that the parameter $s$, which describes the radial offset of the solvent induced ion potentials $V_\pm$
with respect to the dielectric interface at $r=a$, can be positive and negative, depending on the net effect 
of the above-mentioned interfacial effects.
More detailed representations of the interfacial effects are possible at the expense of more phenomenological
parameters \cite{Onuki2006,Onuki2008a,Onuki2008b}, but for the sake of convenience and because handy analytical
expressions are desired the present most simple choice is made here.

A convenient approach to calculate the interfacial tension of the system under consideration is to first
determine the equilibrium ion number density profiles $\rho_\pm$ by means of density functional theory 
\cite{Evans1979,Evans1989,Evans1991} and then to infer the interfacial tension from inserting these equilibrium 
profiles into the grand potential density functional. 
Poisson-Boltzmann theory corresponds to the mean-field grand potential density functional 
\begin{eqnarray}
   \Omega[\rho_\pm] 
   & = & 
   4\pi\sum_{\alpha=\pm}\Int{0}{\infty}{r} r^2\rho_\alpha(r)\Big(\ln(\rho_\alpha(r))-1-\mu_\alpha
   \nonumber\\
   &&
   \phantom{4\pi\sum_{\alpha=\pm}\Int{0}{\infty}{r}}
   +V_\alpha(r)+\frac{\alpha}{2}\phi(r,[\rho_\pm])\Big)
   \label{eq:df}
\end{eqnarray}
with $\mu_\alpha$ the chemical potential of $\alpha$ ions and $\phi(r,[\rho_\pm])$ the electrostatic potential
functional at radius $r$, which fulfills the Poisson equation
\begin{equation}
   \frac{1}{r^2}\Big(\epsilon(r)r^2\phi'(r,[\rho_\pm])\Big)' 
   = 
   -4\pi\sum_{\alpha=\pm}\alpha\rho_\alpha(r)
   \label{eq:pe}
\end{equation}
subject to the boundary conditions $\phi'(r=0)=0$ and $\phi(r=\infty)=0$, where a prime denotes a derivative with
respect to $r$ and $\epsilon(r):=\epsilon_\m{d}\Theta(a-r)+\epsilon_\m{b}\Theta(r-a)$. 
The electrostatic potential is a continuous function of $r$, and at the dielectric interface ($r=a$) the radial 
component of the dielectric displacement is continuous: 
$\epsilon_\m{d}\phi'(r\nearrow a)=\epsilon_\m{b}\phi'(r\searrow a)$.

Minimizing the density functional in Eq.~\Eq{df} gives rise to the Euler-Lagrange equations
\begin{equation}
   \rho_\alpha(r) = \exp(\mu_\alpha - V_\alpha(r) - \alpha\phi(r,[\rho_\pm])).
   \label{eq:ele1}
\end{equation}
Due to the local charge neutrality in the bulk liquid far away from the droplet ($\rho_+(r=\infty)=\rho_-(r=\infty)$)
one infers $\mu_+=\mu_-=:\mu$. 
Upon introducing the reference densities $\rho^\m{ref}_\m{b}:=\exp(\mu)$ and 
$\rho^\m{ref}_\m{d}:=\rho^\m{ref}_\m{b}\exp(-(f_++f_-)/2)$, the sharp-kink reference density profile 
$\rho^\m{ref}(r,x):=\rho^\m{ref}_\m{b}\Theta(r-x) + \rho^\m{ref}_\m{d}\Theta(x-r)$ with the discontinuity located
at radius $x$, and the shifted electrostatic potential $\psi(r):=\phi(r)-\phi_D\Theta(a+s-r)$ with the Donnan 
potential $\phi_D:=(f_--f_+)/2$, the Euler-Lagrange equation~\Eq{ele1} can be rewritten as 
\begin{equation}
   \rho_\alpha(r) = \rho^\m{ref}(r,a+s)\exp(-\alpha\psi(r)).
   \label{eq:ele2}
\end{equation}

Inserting Eq.~\Eq{ele2} into the Poisson equation~\Eq{pe} leads to the Poisson-Boltzmann equation
\begin{equation}
   \psi''(r) = \kappa(r)^2\sinh(\psi(r)), \qquad r\not=a,a+s
   \label{eq:pbe}
\end{equation}
with $\kappa(r):=\sqrt{8\pi\rho^\m{ref}(r,a+s)/\epsilon(r)}$ the Debye screening factor.
Given a solution $\psi$, the interfacial tension with respect to the dielectric interface at $r=a$ in excess to
the pure, salt-free liquid-liquid interfacial tension between the droplet and the bulk liquid is, after 
inserting Eq.~\Eq{ele2} into Eq.~\Eq{df}, determined by
\begin{eqnarray}
   \gamma^\m{ex}
   &\!\! = \!\!&
   \frac{\Omega[\rho_\pm] - \Omega[\rho^\m{ref}(\cdot,a)]}{4\pi a^2}
   \nonumber\\
   &\!\! = \!\!&
   -\frac{1}{a^2}\sum_{\alpha=\pm}\Int{0}{\infty}{r} 
   r^2\Big(\rho_\alpha(r) - \rho^\m{ref}(r,a)
   \nonumber\\
   & &
   \phantom{-\frac{1}{a^2}\sum_{\alpha=\pm}\Int{0}{\infty}{r}}
   +\frac{\alpha}{2}\rho_\alpha(r)\phi(r,[\rho_\pm])\Big).
   \label{eq:gamma1}
\end{eqnarray}
As solutions of the non-linear Poisson-Boltzmann equation~\Eq{pbe} in the spherical geometry can be obtained only
numerically, the same holds for the excess interfacial tension $\gamma^\m{ex}$ in Eq.~\Eq{gamma1}.
However, upon linearizing the Euler-Lagrange equation~\Eq{ele2} and the Poisson-Boltzmann equation~\Eq{pbe} one
obtains analytical expressions for the excess interfacial tension $\gamma^\m{ex}$, which will be derived in the 
next section.


\section{\label{sec:linear}Linearized theory}

For a sufficiently small Donnan potential $|\phi_D| < 1$ the Euler-Lagrange equations~\Eq{ele2} and the 
Poisson-Boltzmann equation~\Eq{pbe} can be linearized leading to
\begin{equation}
   \rho_\alpha(r) = \rho^\m{ref}(r,a+s)(1-\alpha\psi(r))
\end{equation}
and
\begin{equation}
   \psi''(r) = \kappa(r)^2\psi(r), \qquad r\not=a,a+s,
   \label{eq:lpbe}
\end{equation}
respectively. 
Inserting both expressions into Eq.~\Eq{gamma1} one obtains
\begin{eqnarray}
   \gamma^\m{ex} 
   & = &
   2s(\rho^\m{ref}_b - \rho^\m{ref}_d)\Big(1+\frac{s}{a}+\frac{1}{3}\Big(\frac{s}{a}\Big)^2\Big)
   \nonumber\\
   & &
   - \frac{\phi_D}{2}\Big(1+\frac{s}{a}\Big)^2\sigma(a+s)
   \label{eq:gamma2}
\end{eqnarray}
with 
\begin{equation}
   \sigma(r):=-\frac{\epsilon(r)\psi'(r)}{4\pi}
   \label{eq:sigma1}
\end{equation}
the charge enclosed by a sphere of radius $r$ around the origin per sphere surface area.

The linear Poisson-Boltzmann equation~\Eq{lpbe} is analytically soluble which gives rise to an expression of
the form
\begin{equation}
   \sigma(a+s) = \frac{\phi_D}{\dps\Big(1+\frac{s}{a}\Big)^2}\sqrt{\frac{\epsilon_\m{b}\rho^\m{ref}_\m{b}}{2\pi}}
                 F(s/a,\kappa_\m{b}a,n,p),
   \label{eq:sigma2}
\end{equation}
where $\kappa_\m{b}:=\sqrt{8\pi\rho^\m{ref}_\m{b}/\epsilon_\m{b}}$, $n:=\sqrt{\epsilon_\m{d}/\epsilon_\m{b}}$,
and $p:=\sqrt{\rho^\m{ref}_\m{d}/\rho^\m{ref}_\m{b}}$. 
The full scaling function $F$, which is recorded in the appendix, appears somewhat lengthy but is straightforward
to obtain in principle.

However, since the effective interfacial width parameter $s$ is usually very much smaller than the droplet radius 
and the local Debye lengths, $|s|\ll a,\kappa(r)^{-1}$, the first argument of the scaling function $F$ can, within
an excellent approximation, be set to zero. 
Inserting $x=0$ into Eqs.~\Eq{F} and \Eq{T} leads to
\begin{widetext}
\begin{equation}
   F(0,y,n,p) = 
   \frac{\dps np + \frac{n(p-n)}{y} - \Big(\frac{n}{y}\Big)^2 + 
              \exp\Big(-2y\frac{p}{n}\Big)\Big(np + \frac{n(p+n)}{y} + \Big(\frac{n}{y}\Big)^2\Big)}
        {\dps 1 + np + \frac{1-n^2}{y} +
              \exp\Big(-2y\frac{p}{n}\Big)\Big(-1 + np - \frac{1-n^2}{y}\Big)}.
   \label{eq:F0}
\end{equation} 
\end{widetext}
At this level of approximation Eqs.~\Eq{gamma2} and \Eq{sigma2} reduce to
\begin{equation}
   \gamma^\m{ex} = 2s\rho^\m{ref}_\m{b}(1-p^2) - \frac{\phi_D}{2}\sigma(a)
   \label{eq:gamma3}
\end{equation}
and
\begin{equation}
   \sigma(a) = \phi_D\sqrt{\frac{\epsilon_b\rho^\m{ref}_\m{b}}{2\pi}}F(0,\kappa_\m{b}a,n,p),
   \label{eq:sigma3}
\end{equation}
respectively.
According to Eq.~\Eq{sigma3}, the droplet charge per droplet surface area $\sigma(a)$ is (almost) independent
of the interfacial width $s$.
On the other hand, the excess interfacial tension $\gamma^\m{ex}$ in Eq.~\Eq{gamma3} comprises a contribution
describing the ion exclusion due to the short-ranged \emph{interfacial effects}
\begin{equation}
   \gamma^\m{ex}_\m{ie} := 2s\rho^\m{ref}_\m{b}(1-p^2),
   \label{eq:gammaie}
\end{equation}
which is (essentially) linear in $s$ and (almost) independent of the droplet radius $a$, as well as an 
\emph{electrostatic} contribution
\begin{equation}
   \gamma^\m{ex}_\m{es} := - \frac{\phi_D}{2}\sigma(a),
   \label{eq:gammaes}
\end{equation}
which is (almost) independent of the effective interfacial width $s$. 

In order to understand the involved dependence of the scaling function $F(0,y,n,p)$ on $y$ it is useful to
investigate the asymptotic behavior for large and small values of $y$.
If $y \gg y^\times_1 := n/p$, the terms in Eq.~\Eq{F0} proportional to the exponentials may be neglected such that
\begin{equation}
   F(0,y\gg y^\times_1,n,p) \simeq \frac{np}{1+np}G(y,n,p)
   \label{eq:F0lowcurv1}
\end{equation}
with
\begin{equation}
   G(y,n,p) := \frac{\dps \Big(1 -\frac{n}{py}\Big)\Big(1 + \frac{1}{y}\Big)}{\dps 1 + \frac{1-n^2}{(1+np)y}}.
   \label{eq:G1}
\end{equation}
Defining $\dps y^\times_2 := \frac{|1-n^2|}{1+np}$ and $y^\times_3 := 1$ and noting that 
$\dps\frac{1-n^2}{1+np}\in[-y^\times_1,y^\times_3]$ one infers from Eq.~\Eq{F0lowcurv1} the leading order 
asymptotic behavior
\begin{equation}
   F(0,y \gg y^\times_1,n,p)\simeq
   \left\{\begin{array}{ll}
      \dps \frac{np}{1+np}       & \m{(I)}\,   y \gg y^\times_3                \\[8pt]
      \dps \frac{np}{1+np}y^{-1} & \m{(II)}\,  y^\times_2 \ll y \ll y^\times_3 \\[8pt]
      \dps \frac{np}{1-n^2}      & \m{(III)}\, y \ll y^\times_2.
   \end{array}\right.
   \label{eq:F0lowcurv2}
\end{equation}
The three cases considered in Eq.~\Eq{F0lowcurv2} are exhaustive and mutually exclusive for $y \gg y^\times_1$
because $y^\times_2 \leq \max(y^\times_1,y^\times_3)$.
If $y \ll y^\times_1$, Eq.~\Eq{F0} leads to 
\begin{equation}
   F(0,y \ll y^\times_1,n,p) \simeq \frac{p^2}{3}y.
   \label{eq:F0highcurv}
\end{equation}

\begin{figure}[!t]
   \includegraphics[width=8.5cm]{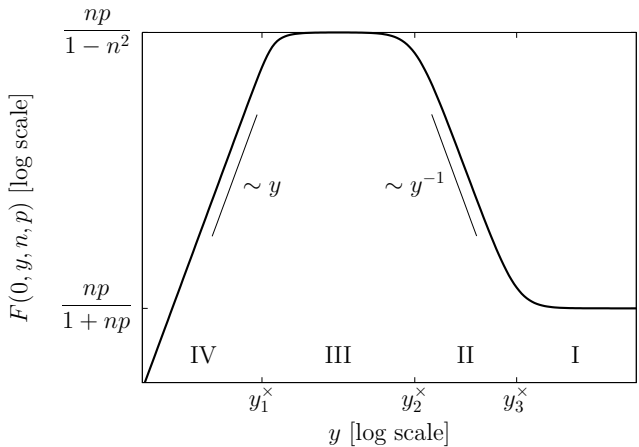}
   \caption{\label{fig:1}Scaling function $F(0,y,n,p)$ as a function of $y$ for the relation
           $y^\times_1 \ll y^\times_2 \ll y^\times_3$ of the crossover positions (see main text) in a log-log plot. 
           The asymptotic regimes I--IV corresponding to Eqs.~\Eq{F0lowcurv2} and \Eq{F0highcurv} are apparent.
           For $y^\times_1 \gg y^\times_2$ regime III is absent, and for $y^\times_1 \gg y^\times_3$ also regime II.}
\end{figure}
Figure~\ref{fig:1} displays $F(0,y,n,p)$ for the case $y^\times_1 \ll y^\times_2 \ll y^\times_3$, where all four
asymptotic regimes I--IV of Eqs.~\Eq{F0lowcurv2} and \Eq{F0highcurv} are apparent. 
If $y^\times_2 \ll y^\times_1 \ll y^\times_3$, however, regime III in Fig.~\ref{fig:1} is absent, and a 
crossover between regimes II and IV takes place at $y=y^\times_1$. 
Moreover, if $y^\times_1 \gg y^\times_3$ regime II is also absent, and $F(0,y,n,p)$ exhibits a single
crossover at $y=y^\times_1$ between regimes I and IV.

It will turn out in the next section that the analytical expressions based on the linearized theory derived in the
present section agree qualitatively, in typical cases even quantitatively, with numerically calculated 
interfacial tensions within the non-linear theory.  


\section{\label{sec:discuss}Discussion}

Here the closed analytical expressions obtained within the linearized Poisson-Boltzmann theory of the previous 
section are discussed and compared with numerical results obtained within the non-linear theory based on 
Eqs.~\Eq{ele2}--\Eq{gamma1}.
Some of the numerical data presented here have already been considered in Ref.~\cite{deGraaf2008}.
Throughout this section one of the liquids is water with dielectric constant $\epsilon_\m{w}=80$.
Moreover, the largely arbitrary but representative choice of ion radii $a_+=0.36\,\m{nm}$ and 
$a_-=0.30\,\m{nm}$ is made throughout.
Given the dielectric constant of the second liquid, called ``oil'', the parameters $n$, $p$, and $\phi_D$ are
known within the Born approximation (see Sec.~\ref{sec:model}).
The cases of an oil droplet in water (O/W) and of a water droplet in oil (W/O) will be distinguished.

\begin{figure}[!t]
   \includegraphics[width=8.5cm]{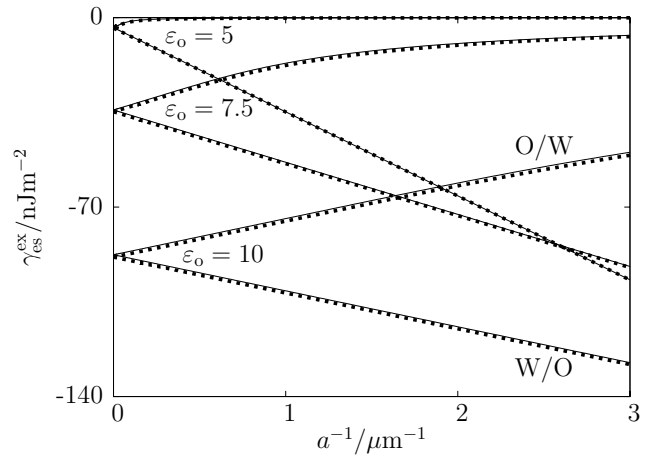}
   \caption{\label{fig:2}Electrostatic contribution to the excess interfacial tension $\gamma^\m{ex}_\m{es}$
           in mixtures of oil ($\epsilon_\m{o}\in\{5,7.5,10\}$) and water ($\epsilon_\m{w}=80$) as a function 
           of the radius $a$ of an oil droplet in water (O/W, ascending curves) and a water droplet in oil (W/O, 
           descending curves) with ion radii $a_+=0.36\,\m{nm}$ and $a_-=0.30\,\m{nm}$ as well as an ionic strength
           in water $I_\m{w}=1\,\m{mM}$. 
           The thin solid curves are calculated by means of the analytical expressions within the linear theory
           of Sec.~\ref{sec:linear} whereas the thick dotted curves are obtained by numerically solving the 
           non-linear Poisson-Boltzmann equation~\Eq{pbe}.
           Upon swapping oil and water (O/W $\leftrightarrow$ W/O) the slope of the curves at $a^{-1}=0$ (planar
           system) changes its sign.}
\end{figure}
\begin{figure}[!t]
   \includegraphics[width=8.5cm]{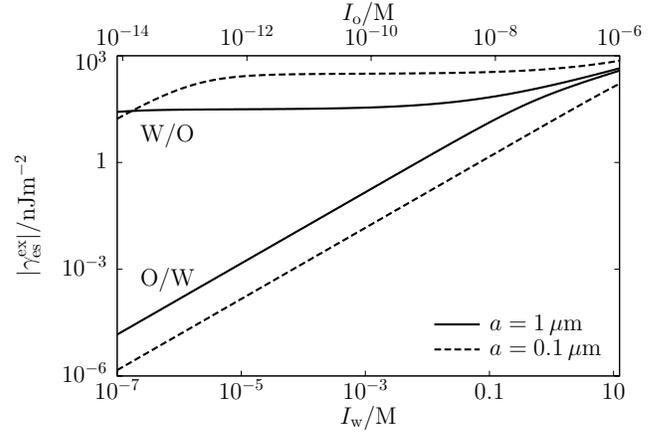}
   \caption{\label{fig:5}Electrostatic excess interfacial tension in mixtures of oil ($\epsilon_\m{o}=5$)
           and water ($\epsilon_\m{w}=80$) as a function of the ionic strength in oil ($I_\m{o}$) or water
           ($I_\m{w}$) for droplet radii $a\in\{0.1\,\m{\mu m},1\,\m{\mu m}\}$
           of an oil droplet in water (O/W) and a water droplet in oil (W/O)
           with ion radii $a_+=0.36\,\m{nm}$ and $a_-=0.30\,\m{nm}$.
           The O/W system exhibits only the regimes I and IV (see main text and Fig.~\ref{fig:1}), whereas
           for the W/O system the regimes I, II, and IV are present.
           Upon changing the droplet size $a$ the crossover ionic strengths shift by a factor $a^{-2}$.}
\end{figure}

\begin{table}[!t]
   \begin{minipage}{8.5cm}
   (a) O/W\\[5pt]
   \begin{tabular}{c||c|c|c|c|c}
      $\epsilon_\m{o}$ & $n$     & $p$        & $\phi_D$ & $y^\times_1$ & $y^\times_2$ \\
      \hline
      $5$              & $0.25$  & $0.000285$ & $1.48$   & $876$        & $0.937$      \\
      $7.5$            & $0.306$ & $0.00520$  & $0.956$  & $58.9$       & $0.905$      \\
      $10$             & $0.354$ & $0.0222$   & $0.692$  & $15.9$       & $0.868$      \\
   \end{tabular}

   \vspace{10pt}

   (b) W/O\\[5pt]
   \begin{tabular}{c||c|c|c|c|c}
      $\epsilon_\m{o}$ & $n$    & $p$    & $\phi_D$ & $y^\times_1$ & $y^\times_2$ \\
      \hline
      $5$              & $4$    & $3500$ & $-1.48$  & $0.00114$    & $0.00107$    \\
      $7.5$            & $3.27$ & $192$  & $-0.956$ & $0.0170$     & $0.0154$     \\
      $10$             & $2.83$ & $45.1$ & $-0.692$ & $0.0627$     & $0.0545$
   \end{tabular}
   \end{minipage}
   \caption{\label{tab:1}Quantities $n$, $p$, and $\phi_D$ as well as the crossover values $y^\times_1$ and 
           $y^\times_2$ (see Secs.~\ref{sec:model} and \ref{sec:linear}) within the Born approximaion for (a) 
           O/W and (b) W/O systems with the oil dielectric constant $\epsilon_\m{o}\in\{5,7.5,10\}$ and
           ion radii $a_+=0.36\,\m{nm}$ and $a_-=0.30\,\m{nm}$.}
\end{table}
Figure~\ref{fig:2} displays the electrostatic contribution to the excess interfacial tension $\gamma^\m{ex}_\m{es}$
(see Eq.~\Eq{gammaes}) of oil droplets in water (O/W, ascending curves) and water droplets in oil (W/O, descending
curves) for an ionic strength in water $I_\m{w}=1\,\m{mM}$, where $I_\m{w}:=\rho^\m{ref}_\m{b}$ for O/W and
$I_\m{w}:=\rho^\m{ref}_\m{d}$ for W/O, as a function of the droplet radius $a$.
The analytical expression Eq.~\Eq{gammaes} within linearized Poisson-Boltzmann theory (thin solid curves) is 
compared with numerical results of the non-linear Poisson-Boltzmann theory (thick dotted curves).
The slight quantitative differences are due to the linearization approximation and they are already present in the
planar system ($a^{-1}=0$).
The quantities $n$, $p$, and $\phi_D$ as well as the crossover values $y^\times_1$ and $y^\times_2$ correponding
to the curves in Fig.~\ref{fig:2} are displayed in Tab.~\ref{tab:1}.
According to Sec.~\ref{sec:linear}, regime III is expected to be absent for the W/O systems because
$y^\times_3 > y^\times_1 > y^\times_2$, whereas regimes II and III are absent for the O/W systems because 
$y^\times_1 > y^\times_3 > y^\times_2$ (see also the discussion of Fig.~\ref{fig:5} at the end of this section).

Due to Eq.~\Eq{gammaes} the relative change of the electrostatic excess 
interfacial tension $\gamma^\m{ex}_\m{es}(a)$ and the droplet charge per droplet surface area $\sigma(a)$ with 
respect to their planar values $\gamma^\m{ex}_\m{es}(\infty)$ and $\sigma(\infty)$, respectively, are equal, and
they exhibit the low-curvature asymptotic behavior (see Eqs.~\Eq{sigma3} and \Eq{F0lowcurv1})
\begin{equation}
   \frac{\gamma^\m{ex}_\m{es}(a)}{\gamma^\m{ex}_\m{es}(\infty)} =
   \frac{\sigma(a)}{\sigma(\infty)} \simeq
   G(\kappa_\m{b}a,n,p)
   \qquad , \kappa_\m{b}a \gg y^\times_1,
   \label{eq:gammasigmalowcurv}
\end{equation}
where $G$ was defined in Eq.~\Eq{G1}.
Upon rewriting Eq.~\Eq{G1} one recognizes the asymptotic behavior
\begin{eqnarray}
   & & G(y \gg y^\times_2,n,p) 
   \nonumber\\
   & = & 
   \dps 1 - \frac{n(1-p^2)}{p(1+np)}y^{-1} 
        - \frac{\dps \frac{n(p+n)^2}{p(1+np)^2}y^{-2}}{\dps 1 + \sign(1-n)\frac{y^\times_2}{y}}
   \nonumber\\
   & \simeq &
   1 - \frac{n(1-p^2)}{p(1+np)}y^{-1} - \frac{n(p+n)^2}{p(1+np)^2}y^{-2},
   \label{eq:G2}
\end{eqnarray}
which equals the expansion in $y^{-1}$ up to second order. 
Hence, the low-curvature expansion up to second order in $a^{-1}$ obtained by combining Eqs.~\Eq{gammasigmalowcurv}
and \Eq{G2} is expected to be acurate if $\kappa_\m{b}a \gg y^\times_1, y^\times_2$.
Traditionally, empirically motivated expansions in $a^{-1}$ have been used to represent the curvature dependence of 
the interfacial tension without knowing their applicability a priori.
However, it has been argued by K\"{o}nig, Roth, and Mecke on the basis of a morphometrical approach that the 
deviation of intensive thermodynamic quantities from their planar values are linear combinations of the
mean and the Gaussian curvature \emph{provided} the geometrical length scales are much larger than any
correlation length \cite{Koenig2004}, i.e., $a \gg \kappa_\m{d}^{-1},\kappa_\m{b}^{-1}$ with the inverse Debye 
length in the droplet $\kappa_\m{d} := \kappa_\m{b}p/n$, or equivalently $\kappa_\m{b}a \gg y^\times_1,y^\times_3$.
This condition is only sufficient but not necessary for the validity of the above low-curvature expansion because
it already implies $\kappa_\m{b}a \gg y^\times_2$ due to $y^\times_2 \leq \max(y^\times_1,y^\times_3)$.
For $n,p \gg 1$, the low-curvature expansion is valid if $\kappa_\m{d}a \gg 1$, \emph{independent} of the bulk 
Debye length $\kappa_\m{b}^{-1}$, because in this case $y^\times_2 \approx y^\times_1$.
This is the case, e.g., for the W/O systems considered in Tab.~\ref{tab:1}.

From Eq.~\Eq{G1} one straightforwardly recognizes the symmetry $G(py/n,1/n,1/p)=G(-y,n,p)$ which means that 
swapping droplet and bulk liquid, i.e., $p \mapsto 1/p$, $n \mapsto 1/n$, $\kappa_\m{b} \mapsto \kappa_\m{d}$, while
keeping the droplet radius $a$ fixed has numerically the same effect on function $G$ as inverting the sign of the
droplet radius.
Due to this symmetry one concludes for the coefficients of an expansion in inverse powers of $a$ as in Eq.~\Eq{G2}
for $y=\kappa_\m{b}a$ that upon swapping droplet and bulk liquid the odd-order coefficients merely invert their 
sign, whereas the even-order coefficients do not change.
This phenomenon can be observed in Fig.~\ref{fig:2}, where the slope close to the planar limit ($a^{-1}=0$), which is
proportional to the excess Tolman length due to the presence of ions, simply changes its sign upon swapping oil 
and water (O/W $\leftrightarrow$ W/O). 

\begin{figure}[!t]
   \includegraphics[width=8.5cm]{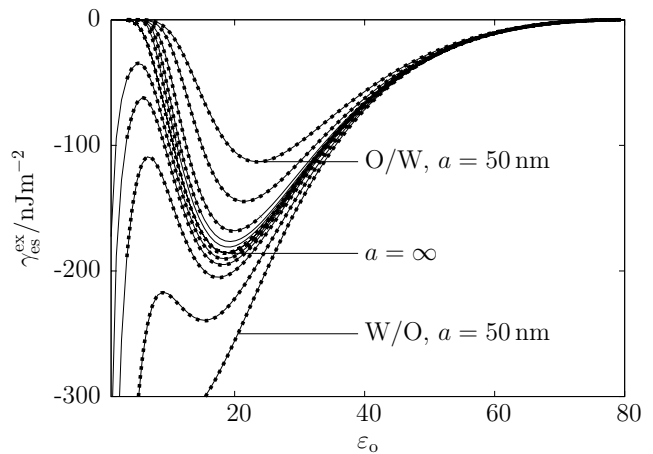}
   \caption{\label{fig:3}Electrostatic contribution to the excess interfacial tension in mixtures of oil 
           and water ($\epsilon_\m{w}=80$) as a function of the dielectric constant $\epsilon_\m{o}$ of the oil
           for droplet radii $a\in\{50\,\m{nm},100\,\m{nm},250\,\m{nm},500\,\m{nm},1000\,\m{nm},\infty\}$
           of an oil droplet in water (O/W) and a water droplet in oil (W/O)
           with ion radii $a_+=0.36\,\m{nm}$ and $a_-=0.30\,\m{nm}$ as well as the ionic strength in water 
           $I_\m{w}=1\,\m{mM}$. 
           The thin solid curves are calculated by means of the analytical expressions within the linear theory
           of Sec.~\ref{sec:linear} whereas the thick dashed curves are obtained by numerically solving the 
           non-linear Poisson-Boltzmann equation~\Eq{pbe}.
           There is quantitative agreement for all $a$ and $\epsilon_\m{o}$ considered here.}
\end{figure}
Figure~\ref{fig:3} exhibits the electrostatic contribution to the excess interfacial tension $\gamma^\m{ex}_\m{es}$
as a function of the dielectric constant $\epsilon_\m{o}$ of the oil for the ionic strength in water 
$I_\m{w}=1\,\m{mM}$ and for various droplet radii 
$a\in\{50\,\m{nm},100\,\m{nm},250\,\m{nm},500\,\m{nm},1000\,\m{nm},\infty\}$.
As in Fig.~\ref{fig:2}, the thin solid curves correspond to the analytic linear theory of Sec.~\ref{sec:linear}
whereas the thick dotted curves are the numerical results of the non-linear scheme.
Quantitative agreement is observed, even in the low-$\epsilon_\m{o}$ range where the Donnan potential $\phi_D$ is 
\emph{not} small and the linearization approximation is \emph{not a priori} justified.
From the linearized theory of Sec.~\ref{sec:linear} one can derive the asymptotic behavior
$\gamma^\m{ex}_\m{es}=\mathcal{O}(-(\epsilon_\m{o}-\epsilon_\m{w})^2)$ for $\epsilon_\m{o}\rightarrow\epsilon_\m{w}$
as well as $\gamma^\m{ex}_\m{es}=\mathcal{O}(-\exp(-\m{const}/\epsilon_\m{o}))$ for an O/W system and 
$\gamma^\m{ex}_\m{es}=\mathcal{O}(-1/\epsilon_\m{o})$ for a W/O system as $\epsilon_\m{o} \rightarrow 0$.
This behavior is apparent in Fig.~\ref{fig:3}, too.

The total excess interfacial tension $\gamma^\m{ex}$ comprises not only the electrostatic part 
$\gamma^\m{ex}_\m{es}$ but also the contribution $\gamma^\m{ex}_\m{ie}$ due to the interfacial effects (see 
Eq.~\Eq{gammaie}).
It is readily seen that $\gamma^\m{ex}_\m{ie}=\pm\mathcal{O}(s(\epsilon_\m{w}-\epsilon_\m{o}))$ for 
$\epsilon_\m{o}\rightarrow\epsilon_\m{w}$ and $\gamma^\m{ex}_\m{ie}=\pm\mathcal{O}(s)$ for 
$\epsilon_\m{o} \rightarrow 0$ where the upper ($+$) and the lower ($-$) sign correspond to an O/W and a W/O 
system, respectively.
Hence, if $s \not= 0$, the interfacial effects will dominate over the electrostatic effects in the limits 
$\epsilon_\m{o} \rightarrow 0$ for O/W systems and $\epsilon_\m{o}\rightarrow\epsilon_\m{w}$ for arbitrary systems.
\begin{figure}[!t]
   \includegraphics[width=8.5cm]{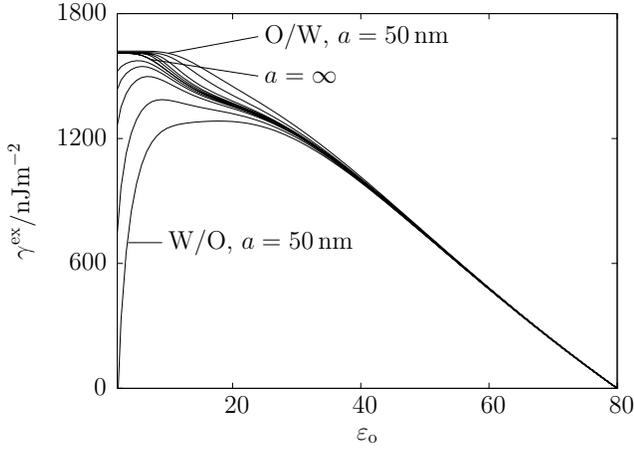}
   \caption{\label{fig:4}Total excess interfacial tension in mixtures of oil 
           and water ($\epsilon_\m{w}=80$) as a function of the dielectric constant $\epsilon_\m{o}$ of the oil
           for droplet radii $a\in\{50\,\m{nm},100\,\m{nm},250\,\m{nm},500\,\m{nm},1000\,\m{nm},\infty\}$
           of an oil droplet in water (O/W) and a water droplet in oil (W/O)
           with ion radii $a_+=0.36\,\m{nm}$ and $a_-=0.30\,\m{nm}$, interfacial width parameter $|s|=0.33\,\m{nm}$,
           as well as the ionic strength in water $I_\m{w}=1\,\m{mM}$.}
\end{figure}
Figure~\ref{fig:4} displays the total excess interfacial tension corresponding to the parameters used in 
Fig.~\ref{fig:3} and an interfacial width parameter $s$ with $|s|=0.33\,\m{nm}$ on the water side of the interface,
i.e., $s>0$ for O/W and $s<0$ for W/O.

According to the results of Sec.~\ref{sec:linear} the electrostatic excess interfacial tension $\gamma^\m{ex}_\m{es}$
as a function of the bulk ionic strength $I_\m{b}:=\rho^\m{ref}_\m{b}$ can be asymptotically described by
\begin{equation}
   \gamma^\m{ex}_\m{es} \simeq 
   \left\{\begin{array}{ll}
      \dps -\phi_D^2\sqrt{\frac{\epsilon_\m{b}}{8\pi}}\frac{np}{1+np}I_\m{b}^{1/2}    
      & 
      \m{(I)}\,   I_\m{b} \gg I^\times_{\m{b}1},I^\times_{\m{b}3} 
      \\[10pt]
      \dps -\phi_D^2\frac{\epsilon_\m{b}}{8\pi}\frac{np}{(1+np)a}                
      &
      \m{(II)}\,  I^\times_{\m{b}1},I^\times_{\m{b}2} \ll I_\m{b} \ll I^\times_{\m{b}3} 
       \\[10pt]
      \dps -\phi_D^2\sqrt{\frac{\epsilon_\m{b}}{8\pi}}\frac{np}{1-n^2}I_\m{b}^{1/2} 
      &
      \m{(III)}\, I^\times_{\m{b}1} \ll I_\m{b} \ll I^\times_{\m{b}2}
      \\[10pt]
      \dps -\phi_D^2\frac{p^2}{3}aI_\m{b}
      & 
      \m{(IV)}\,  I_\m{b} \ll I^\times_{\m{b}1}
   \end{array}\right.
   \label{eq:gammaesasymp}
\end{equation}
with the crossover bulk ionic strengths $\dps I^\times_{\m{b}k} := \frac{\epsilon_\m{b}(y^\times_k)^2}{8\pi a^2}, 
k\in\{1,2,3\}$, where the $y^\times_k, k\in\{1,2,3\}$ are defined in Sec.~\ref{sec:linear}.
For a planar system ($a=\infty$) the crossovers are at zero ionic strength, hence only the high-ionic strength 
regime I in Fig.~\ref{fig:1} ($I_\m{b} \gg I^\times_{\m{b}1},I^\times_{\m{b}3}$) is present, which coincides 
exactly with the electrostatic contribution to the excess interfacial tension in Ref.~\cite{Bier2008}.

For an oil dielectric constant $\epsilon_\m{o}=5$ and a droplet radius $a=1\,\m{\mu m}$ the crossover bulk ionic
strengths are $I^\times_{\m{b}1} \approx 71\,\m{mM}, I^\times_{\m{b}2} \approx 82\,\m{nM}, I^\times_{\m{b}3} 
\approx 93\,\m{nM}$ for an O/W system, where $I_\m{b}=I_\m{w}$ is the ionic strength in water, and 
$I^\times_{\m{b}1} \approx 7.6\,\m{fM}, I^\times_{\m{b}2} \approx 6.6\,\m{fM}, I^\times_{\m{b}3} \approx 
5.8\,\m{nM}$ for a W/O system, where $I_\m{b}=I_\m{o}$ is the ionic strength in oil (see Tab.~\ref{tab:1}).
Here, ionic strengths in oil, $I_\m{o}$, and in water, $I_\m{w}$, are related to each other by $I_\m{o}/I_\m{w}
\approx 8.1\cdot 10^{-8}$.
Figure~\ref{fig:5} displays $\gamma^\m{ex}_\m{es}$ as a function of the ionic strength in the physical
range $I_\m{w}\in[10^{-7}\,\m{M},10\,\m{M}]$ for the droplet radii $a=1\,\m{\mu m}$ and $a=0.1\,\m{\mu m}$.
The crossover ionic strengths of the latter droplet size are larger by a factor $100$ as compared to the former
because $I^\times_{\m{b}k} \sim a^{-2}$.
By inspection of the values of the crossover bulk ionic strengths one expects only the regimes I and IV of 
Fig.~\ref{fig:1} to be present for the O/W system whereas the regimes I, II, and IV are expected for the W/O 
system.
The occurrence of the regimes I and IV for the O/W system and I, II, and IV for the W/O system can be
inferred from Fig.~\ref{fig:5} in conjunction with Eq.~\Eq{gammaesasymp}.


\section{\label{sec:conclusions}Conclusions and Summary}

It turned out in the previous section that the analytical theory of Sec.~\ref{sec:linear} based on a linearized
Poisson-Boltzmann theory is in good (at least) qualitative agreement with the results from the full non-linear 
theory.
It can therefore be expected that the general conclusions drawn from that linear theory apply to more
elaborate models \cite{Onuki2006,Onuki2008a,Onuki2008b}, too.

According to Eqs~\Eq{gamma3}, \Eq{gammaie}, and \Eq{gammaes} the excess liquid-liquid interfacial tension is
$\gamma^\m{ex}=\gamma^\m{ex}_\m{ie}+\gamma^\m{ex}_\m{es}$ where the curvature dependence is
essentially only due to the electrostatic part $\gamma^\m{ex}_\m{es}$ and not due to the contribution of the
short-ranged interfacial effects $\gamma^\m{ex}_\m{ie}$.
While $\gamma^\m{ex}$ can indeed be negative, thereby decreasing the total interfacial tension, the largest 
magnitude $|\gamma^\m{ex}|$ is attained at high ionic strengths where $\gamma^\m{ex}\approx\gamma^\m{ex}_\m{ie}$, 
i.e., where $\gamma^\m{ex}$ is essentially curvature-independent.
One has to conclude that the unimodal droplet size distribution of W/O emulsions observed by Leunissen
et al.\ \cite{Leunissen2007a,Leunissen2007b} \emph{cannot} be explained by the curvature dependence of the 
interfacial tension due to electrostatic effects alone.
However, this conclusion does not apply to the experiments by Sacanna et al.\ \cite{Sacanna2007}, where 
highly charged colloids instead of monovalent ions are present, as the linearized theory of 
Sec.~\ref{sec:linear} is not a priori justified for multivalent ions or highly charged colloids.
Instead it is an interesting open question to be addressed in future studies as to what extent the qualitative
low-valency picture drawn here is valid for the presence of high-valency particles.

In summary, the curvature dependence of the electrolytic liquid-liquid interfacial tension within a simple linear 
Poisson-Boltzmann model in the spherical geometry has been calculated analytically.
This linear theory turned out to be at least qualitatively reliable as has been checked by numerically solving
the corresponding non-linear Poisson-Boltzmann model.
Novel low ionic strength regimes, which are not present for a planar liquid-liquid interface, have been identified.
Low and high curvature asymptotics of the interfacial tension have been discussed.
In particular, it has been found that in systems where the ionic strength and the dielectric constant in the droplet
are much larger than in the bulk the range of validity of low-curvature expansions up to second order in the
inverse radius of curvature is independent of the bulk Debye length.


\begin{acknowledgments}
   This work is part of the research program of the ``Stichting voor 
   Fundamenteel Onderzoek der Materie (FOM)'', which is financially supported by 
   the ``Nederlandse Organisatie voor Wetenschappelijk Onderzoek (NWO)''. 
\end{acknowledgments}


\appendix*
\section{Scaling function $F$}

Upon solving the linearized Poisson-Boltzmann equation~\Eq{lpbe} one obtains analytical solutions for the
shifted electrostatic potential $\psi$, which, via Eq.~\Eq{sigma1}, determines the scaling function $F$ 
introduced in Eq.~\Eq{sigma2}:
\begin{widetext}
\begin{equation}
   F(x,y,n,p) =
   \left\{\begin{array}{ll}
      \dps 
      \frac{\dps T_1\Big(p(1+x)-\frac{n}{y}+\exp\Big(-2y\frac{p}{n}(1+x)\Big)\Big(p(1+x)+\frac{n}{y}\Big)\Big)}
           {\dps T_2 + \exp\Big(-2y\frac{p}{n}(1+x)\Big)T_3} 
      & , x<0 \\[20pt]
      \dps 
      p\Big(1+x+\frac{1}{y}\Big)
      \frac{\dps T_4 + \exp\Big(-2y\frac{p}{n}\Big)T_5}
      {\dps T_6 + \exp\Big(-2y\frac{p}{n}\Big)T_7}
      & , x>0,
   \end{array}\right.
   \label{eq:F}
\end{equation}
where
\begin{eqnarray}
   T_1 & := & \dps n\Big(1 + x + \frac{1}{y} + \frac{(1-n^2)x}{y}\Big)\cosh\Big(\frac{xy}{n}\Big) - 
                   n^2\Big(1 + x + \frac{1}{y} + \frac{1-n^2}{y^2}\Big)\sinh\Big(\frac{xy}{n}\Big)
   \nonumber\\
   T_2 & := & \dps \Big(1 + np + \frac{1-n^2}{y}\Big)\cosh\Big(\frac{xy}{n}\Big) - 
                   \Big(n + p + \frac{p(1-n^2)}{y}\Big)\sinh\Big(\frac{xy}{n}\Big)
   \nonumber\\
   T_3 & := & \dps \Big(-1 + np - \frac{1-n^2}{y}\Big)\cosh\Big(\frac{xy}{n}\Big) + 
                   \Big(n - p - \frac{p(1-n^2)}{y}\Big)\sinh\Big(\frac{xy}{n}\Big)
   \nonumber\\
   T_4 & := & \dps \Big(n(1+x) - \frac{n^2}{py} + \frac{(1-n^2)x}{py}\Big)\cosh(pxy) +
                   \Big(1 - \frac{n}{py} + x - \frac{1-n^2}{(py)^2}\Big)\sinh(pxy)
   \nonumber
\end{eqnarray}
\begin{eqnarray}
   T_5 & := & \dps \Big(n(1+x) + \frac{n^2}{py} - \frac{(1-n^2)x}{py}\Big)\cosh(pxy) -
                   \Big(1 + x + \frac{n}{py} - \frac{1-n^2}{(py)^2}\Big)\sinh(pxy)
   \nonumber\\
   T_6 & := & \dps \Big(1 + np + \frac{p(1-n^2)}{py}\Big)\cosh(pxy) +
                   \Big(n + p + \frac{1-n^2}{py}\Big)\sinh(pxy)
   \nonumber\\
   T_7 & := & \dps \Big(np - 1 - \frac{p(1-n^2)}{py}\Big)\cosh(pxy) +
                   \Big(n - p - \frac{1-n^2}{py}\Big)\sinh(pxy).
   \label{eq:T}
\end{eqnarray}
Note that $F(x,y,n,p)$ is continuous at $x=0$.
\end{widetext}



\end{document}